\documentclass[twocolumn,showpacs,preprintnumbers,amsmath,amssymb]{revtex4}
\usepackage{graphicx}
\usepackage{dcolumn}
\usepackage{bm}
\usepackage{epsfig}
\usepackage{bm}

\newcommand{\be}{\begin{equation}}
\newcommand{\ee}{\end{equation}}
\newcommand{\beqq}{\setlength\arraycolsep{2pt}\begin{eqnarray}}
\newcommand{\eeqq}{\vspace{0cm} \end{eqnarray}}
\newcommand{\bea}{\begin{eqnarray}}
\newcommand{\eea}{\end{eqnarray}}

\begin{document}

\title{Are Galaxy Clusters Suggesting an Accelerating Universe Independent of SNe Ia \\and Gravity Metric Theory? }

\author{J. A. S. Lima}\email{limajas@astro.iag.usp.br}
\author{R. F. L. Holanda}\email{holanda@astro.iag.usp.br}
\author{J. V. Cunha} \email{cunhajv@astro.iag.usp.br}

\affiliation{Departamento de Astronomia, Universidade de S\~{a}o
Paulo \\ Rua do Mat\~ao, 1226 - 05508-900, S\~ao Paulo, SP, Brazil}


\bigskip
\begin{abstract}
A kinematic method to access cosmic acceleration based exclusively
on the Sunyaev-Zel'dovich effect (SZE) and X-ray surface brightness
data from galaxy clusters is proposed. By using the SZE/X-ray data
from 38 galaxy clusters in the redshift range $0.14\leq z \leq 0.89$
[Bonament et al., Astrop. J. {\bf 647}, 25 (2006)], we find that
the present Universe is accelerating and that the transition from an
earlier decelerating  to a late time accelerating regime occurred
relatively recent. Such results are fully independent on the
validity of any metric gravity  theory, the possible matter-energy
contents filling the Universe, as well as on the SNe type Ia Hubble
diagram from which the present acceleration was inferred.
The ability of the ongoing Planck satellite mission to obtain
tighter constraints on the expansion history through SZE/X-ray
angular diameters is also discussed. Two simple simulations of future Planck data 
suggest that such technique will be competitive with supernova data besides being
complementary to it.

\end{abstract}

\maketitle


\emph{Introduction.} The dimming of distant type Ia supernovae
observed by two different group of astronomers one decade ago lead
to unexpected and landmark conclusion: the universal expansion is
speeding up and not slowing down as believed since the early days of
observational cosmology\cite{riess98, perl99}.

Such phenomenon is normally interpreted as a dynamic influence of
some sort of dark energy whose main effect is to change the sign of
the decelerating parameter $q(z)$\cite{Padm03}. Another
possibility is that the cosmic acceleration is a manifestation of
new gravitational physics (rather than dark energy) that involves a
modification of the left hand side (geometric sector) of the
Einstein field equations. In this sort of theory the Friedmann
equation is modified and a late time  accelerating stage is obtained  even for a Universe 
filled only with cold dark
matter (CDM)\cite{FR}. At present, the space parameter associated
with the cosmic expansion is too degenerate, and, as such, it is not
possible to decode which mechanism or dark energy component is
operating in the cosmic dynamics\cite{Padm03,FR}.

Currently, SNe type Ia are not only the powerful standard candles
available  but still provides a unique direct access to the late
time accelerating stage  of the Universe. Naturally, this a rather
uncomfortable situation from the observational and theoretical
viewpoints even considering that ten years later, the main
observational concerns about errors in SNe type Ia measurements,
like host galaxy extinction, intrinsic evolution, possible selection
bias in the low redshift sample seem to be  under
control\cite{Kowalski08}.

\begin{figure*}[th]
\centerline{\psfig{figure=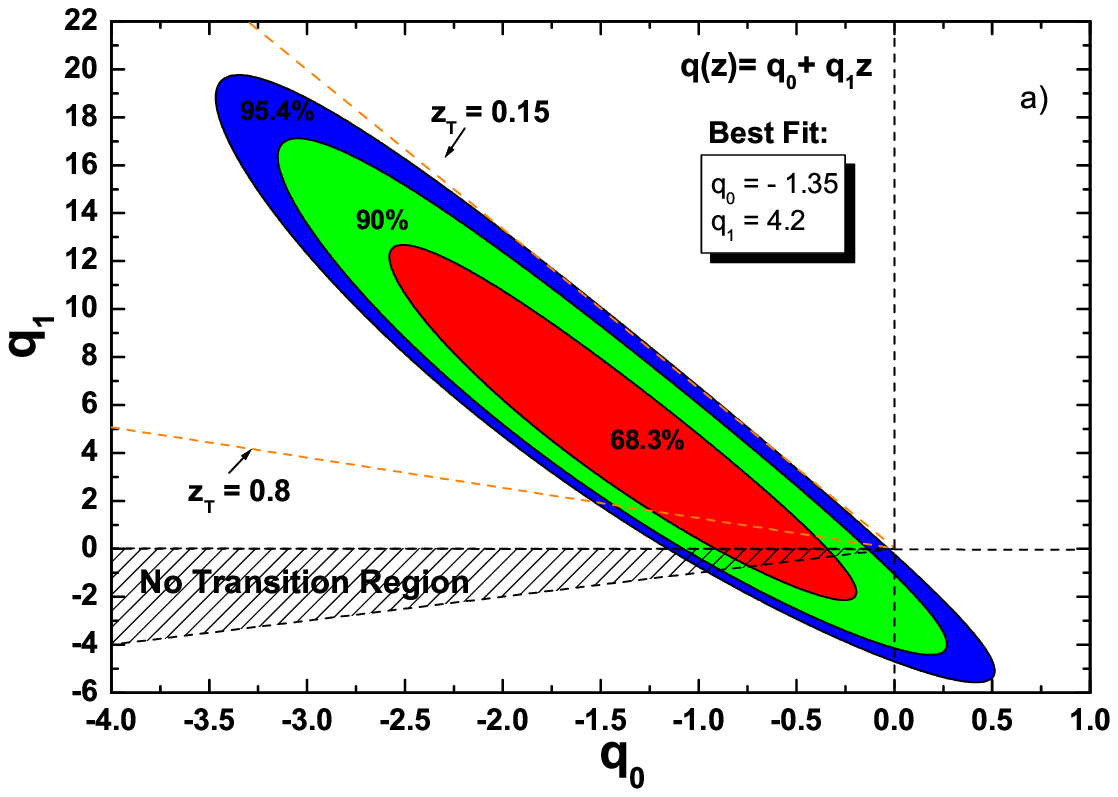,width=3.0truein,height=2.5truein}
\psfig{figure=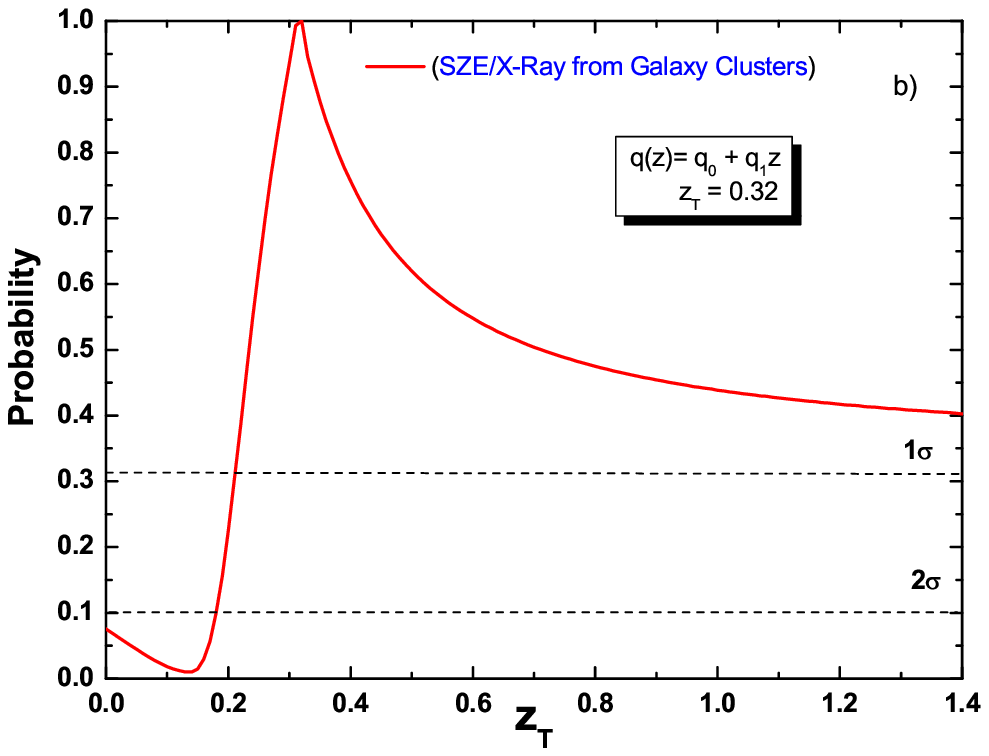,width=3.0truein,height=2.5truein}}
\vspace{-0.7cm}
\caption{ {\bf{a)}} Contours in the $q_o - q_1$ plane for 38 galaxy
clusters data \cite{Boname06} considering $q(z)=q_0 +q_1z$. The best
fit to the pair of free parameters is ($q_0,q_1) \equiv
(-1.35,4.2)$. For comparison we have shown the straight lines
denoting the transitions redshifts for two different $\Lambda$CDM
models: $z_t=0.9$ for $\Omega_{\Lambda}=0.8$ and $z_t=0.15$ for
$\Omega_{\Lambda}=0.43$. {\bf{b)}} Likelihood function for the
transition redshift. The best fit is $z_t = 0.32$.}\label{fig1}
\end{figure*}

A promising estimator fully independent of SNe type Ia and other
calibrators of the cosmic distance ladder is the angular diameter
distance ($D_{A} (z)$) from a given set of distant objects. It has
also been recognized that the combination of SZE \cite{sunzel70} and
X-ray surface brightness measurements  may provide useful angular
diameters from  galaxy clusters
\cite{caval77,Birk99,SZEPapers2,Boname06,CML07}.

On the other hand, since the mechanism causing the acceleration is
still unknown, it is interesting to investigate the potentialities
of SZE/X-ray technique from a more general viewpoint, that is,
independent of the gravity theory and the matter-energy contents
filling the Universe. The better strategy available so far is to
consider the same kind of kinematic approach which has been
successfully applied for determining the transition
deceleration/acceleration in the past by using SNe type Ia
measurements\cite{TR02,Riess04,EM06,CL08,Cunha09}.

In this letter, we employ a purely kinematic description of the
universal expansion based on angular diameter distances of clusters
for two different expansions of the deceleration parameter.  As we
shall see, by using the Bonamente {\it et al.}\cite{Boname06} sample
we find that a kinematic analysis based uniquely on cluster data
suggests that the Universe undergone a ``dynamic phase transition"
(deceleration/acceleration) in a redshift $z \approx 0.3$. Further,
it is also shown that the Planck satellite mission data must provide
very restrictive limits on the space parameter, thereby opening an
alternative route for accessing the expansion history of the
Universe.

\emph{Angular Diameter and Kinematic Approach.} Let us now assume
that the Universe is spatially flat as motivated by inflation and
WMAP measurements \cite{Komat08}. In this case, the angular diameter
distance in the FRW  metric is defined by (in our units $c=1$),
\begin{eqnarray}\label{eq:dLq}
D_A &=& (1+z)^{-1}H^{-1}_{0}\int_0^z {du\over H(u)} = \frac{(1+z)^{-1}}{H_0} \nonumber \\
&& \,\, \int_0^z\, \exp{\left[-\int_0^u\, [1+q(u)]d\ln
(1+u)\right]}\, du,
\end{eqnarray}
where  $H(z)=\dot a/a$ is the Hubble parameter, and, $q(z)$, the
deceleration parameter, is defined by

\begin{eqnarray}\label{qz}
q(z)\equiv -\frac{a\ddot a}{\dot a^2} = \frac{d H^{-1}(z)}{ dt} -1.
\end{eqnarray}

In the framework of a flat FRW metric, Eq. (1) is an exact
expression for the angular diameter distance. As one may check, in
the case of a linear two-parameter expansion for $q(z)=q_0+z{q_1}$,
the above integral can  analytically be represented as

\begin{eqnarray}\label{eq:dAKin}
D_A(z) &=& \frac{(1+z)^{-1}}{H_0}e^{q_1}q_1{^{q_0-q_1}}
[\gamma{({q_1-q_0},(z+1){q_1})} \nonumber \\
&& \,\, - \gamma{({q_1-q_0},{q_1})}],
\end{eqnarray}
where ${q_0}$ and ${q_1}$ are the values of $q(z)$ and its redshift
derivative, $dq/dz$ evaluated at $z=0$ while $\gamma$ is an
incomplete gamma \cite{AbrSte72}. By using the above expressions we
may get information about $q_0$, $q_1$ and, therefore, about the
global behavior of $q(z)$. In principle,  a dynamic ``phase
transition" (from decelerating to accelerating) happens at
$q(z_t)=0$, or equivalently, $z_t=-q_0/q_1$. Another interesting
parametrization  is $q(z)=q_o + q_1 z/(1+z)$ \cite{CL08,Cunha09}. It
has the advantage to be well behaved at high redshift while the
linear approach diverges at the distant past. Now, the integral (1)
assumes the form:
\begin{eqnarray}\label{eq:dAKin2Parame}
D_A (z) &=& \frac{(1+z)^{-1}}{H_{0}}e^{q_{1}}{q_1^{-(q_0+q_1)}}
[\gamma(q_{1}+q_{0},q_1) \nonumber \\
& & \,\, - \gamma{({q_1+q_0},q_1/(1+z))}],
\end{eqnarray}
where  ${q_1}$ now is the parameter yielding the total correction in
the distant past ($z\gg 0, q(z)= q_0 + q_1$) and $\gamma$ is again
the incomplete gamma function.  Note that in this case the
transition redshift is defined by $z_t=-q_0/(q_0+q_1)$.
\begin{figure*}[th]
\centerline{\psfig{figure=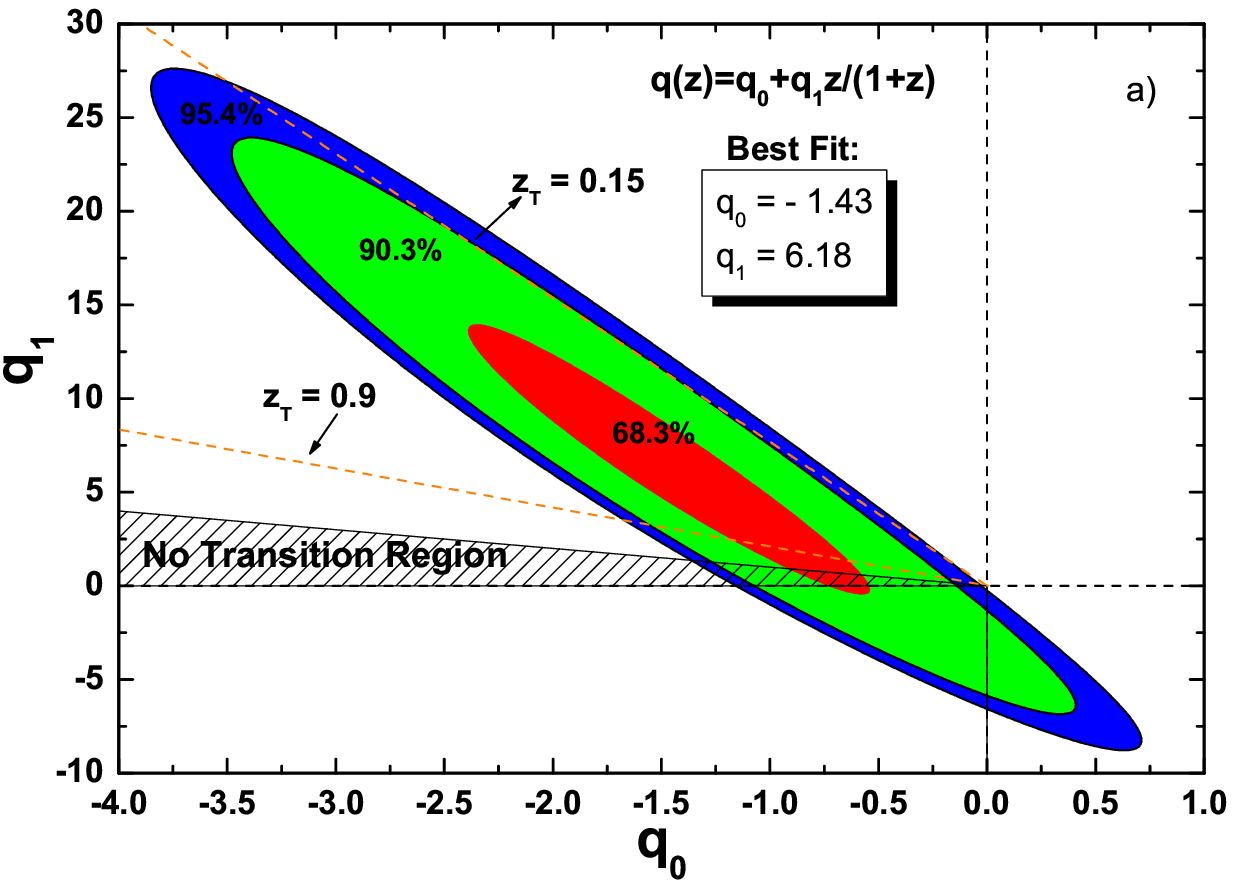,width=3.0truein,height=2.5truein}
\psfig{figure=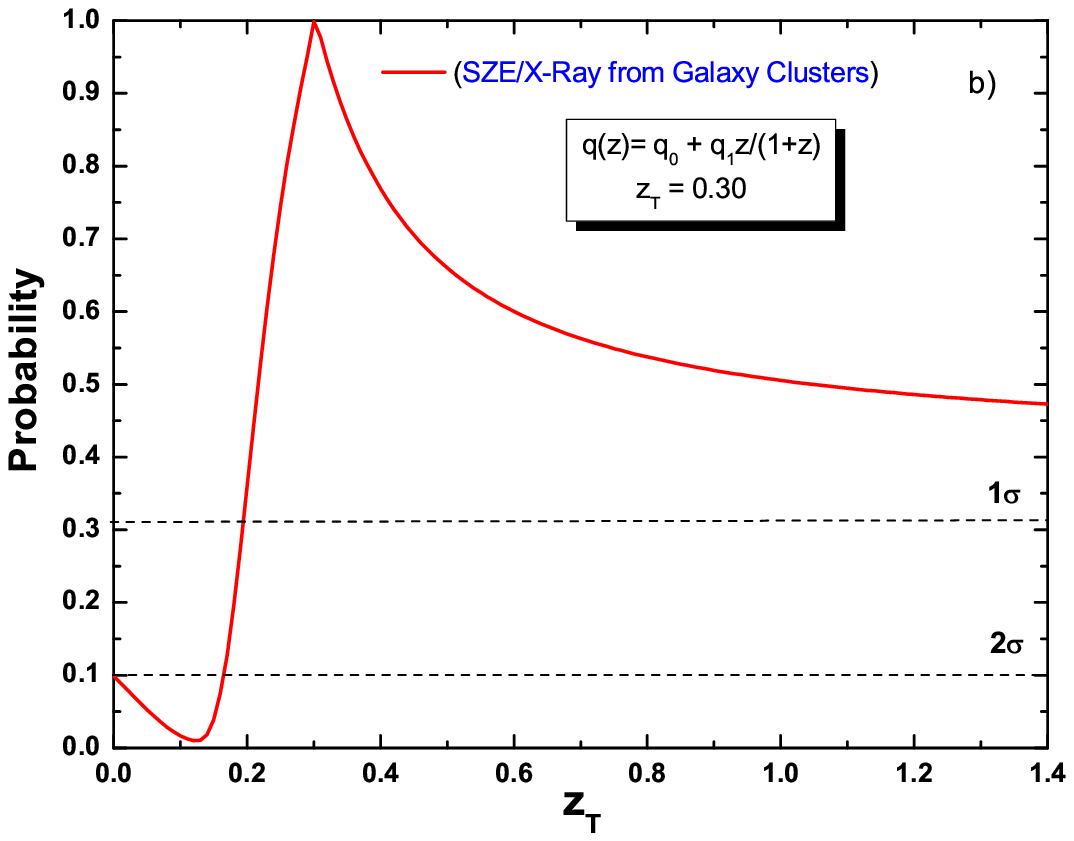,width=3.0truein,height=2.5truein} \hskip
0.1in} \vspace{-0.5cm}
q
c
p
f
\caption{ {\bf{a)}} Contours in the $q_0 - q_1$ plane for 38 galaxy
clusters data \cite{Boname06} considering $q(z)=q_0+q_1z/(1+z)$. The
best fit to the pair of free parameters is ($q_0,q_1$) =
($-1.43,6.18)$. {\bf{b)}} Likelihood function for the transition
redshift and the associated best fit at $z_t=0.30$. Comparing with
Figs. 1(a) and 1(b)  we see that the results are weekly dependent on
the parameterizations.}\label{fig1}
\end{figure*}

\emph{Constraints from Galaxy Clusters.} The SZE is a small
distortion on the Cosmic Microwave Background (CMB) spectrum
provoked by the inverse Compton scattering of the CMB photons
passing through a population of hot electrons. Observing the
temperature  decrement of the CMB spectrum towards galaxy clusters
together the X-rays observations, it is possible to break the
degeneracy between concentration and temperature thereby obtaining
$D_{A}(z)$. Therefore, such distances are fully independent of the
one given by the luminosity distance, $D_L(z)$.

Let us now consider the 38 measurements of angular diameter
distances  from galaxy clusters as obtained through SZE/X-ray method
by Bonamente and coworkers \cite{Boname06}. In our analysis we use a
maximum likelihood determined by a $\chi^{2}$ statistics

\begin{equation}
\chi^2(z|\mathbf{p}) = \sum_i { ({\cal{D}}_A(z_i; \mathbf{p})-
{\cal{D}}_{Ao,i})^2 \over \sigma_{{\cal{D}}_{Ao,i}}^2 +
\sigma_{stat}^{2}}, \label{chi2}
\end{equation}
where ${\cal{D}}_{Ao,i}$ is the observational angular diameter
distance, $\sigma_{{\cal{D}}_{Ao,i}}$ is the uncertainty in the
individual distance, $\sigma_{stat}$ is the contribution of the
statistical errors added in quadrature ($\approx 20$\%) and the
complete set of parameters is given by $\mathbf{p} \equiv (H_{0},
q_{0},q_{1})$. For the sake consistency, the Hubble parameter
$H_{0}$  has been fixed by its best fit value $H^{*}_{0}=80
km/s/Mpc$.

\emph{Linear Parameterization: $q = qo + q_{1}z$.} In Figs. 1(a) and
1(b) we show, respectively,  the contour in the plane $q_{0}-q_{1}$
($68.3\%$, $90\%$ and $95.4\%$ c.l.) and likelihood of the
transition redshift from the Bonamente et al. sample. The confidence
region (1$\sigma$) are $-2.6 \leq q_0 \leq -0.25$ and $13 \leq q_1
\leq -3$. Such results favor a Universe with recent acceleration
($q_0 < 0$) and a previous decelerating stage ($dq/dz > 0$). From
Fig. 1(a) we see that he best fits to the free parameters are $q_{0}
= - 1.35, q_{1} = 4.2$ while for the transition redshift is
$z_{t}=0.32$ (see Fig. 1(b)).  Note the presence of a forbidden
region forming a trapezium. The horizontal line at the top is
defined by $q_1 = 0$, which leads to an infinite (positive or
negative) transition redshift. Note also that the segment at $45\%$
defines the infinite future ($z_t = -1$). In addition, one may
conclude that the vertical segment on the left closing the trapezium
is also unphysical since it is associated $z_t \leq -1$,  thereby
demonstrating that the hatched trapezium is actually a physically
forbidden region (for a similar analysis involving luminosity
distance see \cite{CL08}). For comparison we have also indicated in
Fig. 1(a) the transition redshifts  $z_t= 0.15$ corresponding to a
flat $\Lambda$CDM with $\Omega_{\Lambda} = 0.43$, as well as, $z_t=
0.9$  corresponding to $\Omega_{\Lambda}\simeq 0.8$.

\emph{2nd Parameterization: $q = qo + q_{1}z/1 + z$.} In Figures
2(a) and 2(b) we display the corresponding plots for the second
parameterization. The confidence region (1$\sigma$) is now defined
by: $-2.4 \leq q_0 \leq -0.5$ and $13.5 \leq q_1 \leq 0$. Such
results also favor a Universe with recent acceleration ($q_0 < 0$)
and a previous decelerating stage ($dq/dz > 0$). As indicated in
Fig. 2(a), the best fits to the free parameters are $q_{0} = - 1.43$
and $q_{1} = 6.18$ while for the transition redshift is a little
smaller $z_{t}=0.3$ (see Fig. 1(b)). It should be noticed the
presence of the forbidden region (trapezium) with a minor difference
in comparison with Fig. 1(a), namely, as an effect of the
parameterization, the horizontal line now is at the bottom. Note
also that a decelerating Universe today ($q_0>0$) is only marginally
compatible at $2\sigma$ of statistical confidence.

The results in the $q_{0}-q_{1}$ planes  for both cases suggest
that: (i) the Universe had  an earlier decelerating stage
$(q_{1}=dq/dz
>0)$, and (ii) the Universe has been accelerating ($q_{0}<0$) since
$z\sim 0.3$. A similar result has been previously obtained using SNe
type Ia as standard candles by Shapiro and Turner\cite{TR02}.

\emph{Prospects for Planck Satellite Mission.} Let us discuss
the potentiality of the SZE/X-ray technique when future data from
Planck satellite mission become available\cite{planck}. The
mission is a project from European Space Agency whose
frequency channels were carefully chosen for measurements of thermal
Sunyaev-Zeldovich effect. In principle, the Planck satellite will
see (through SZE) about 30,000 galaxy clusters over the whole sky
with a significant fraction of clusters near or beyond redshift unity.
However, since accurate angular diameter measurements require long
SZE/X-ray integrations nobody expects that all observed clusters
might have useful distance measurements to constrain cosmological
parameters. Therefore, it is interesting to simulate two realistic
samples of angular diameter distances (ADD)  by using a fiducial
model to $D^{true}_{Ao,i}=D_{A}(z_{i}, q^{*}_{0}, q^{*}_{1},
H^{*}_{0})$, where $H^{*}_{0}$, $q^{*}_{0}$ and $q^{*}_{1}$ are the
best fit values to the linear case obtained from Bonamente {\it et
al.} sample\cite{Boname06}.

\begin{center}
\centerline{Table 1} \vspace{0.05cm}

  \begin{tabular}{c@{\quad}c@{\quad}c@{\quad}c@{\quad}c}
    $z$ range &  Clusters  &  bins  &  Clusters/bin  & $ADD_{i}$  Error \\
    & (P, O)& &(P,O) & (P,O) \\ \hline
    $[0.0, 0.5]$ & 100, 500 & 10 & 10, 50 & 15\%, 10\% \\
    $[0.5, 1.0]$ &  70, 350 & 10 & 7, 35    & 17\%, 12\% \\
    $[1.0, 1.5]$ &  40, 200 & 10 & 4, 20    & 20\%, 15\% \\

 \end{tabular}
\end{center}

The first simulation (termed pessimistic - P), assumes that only 210
clusters are distributed in the redshift ranges in the following
form: $0\leq z \leq 0.5$ (100), $0.5\leq z \leq 1$ (70) and $1\leq z
\leq 1.5$ (40) with ADD statistical errors of $15\%$, $17\%$ and
$20\%$, respectively (see Table 1).

\begin{figure}[h!]
\centerline{\psfig{figure=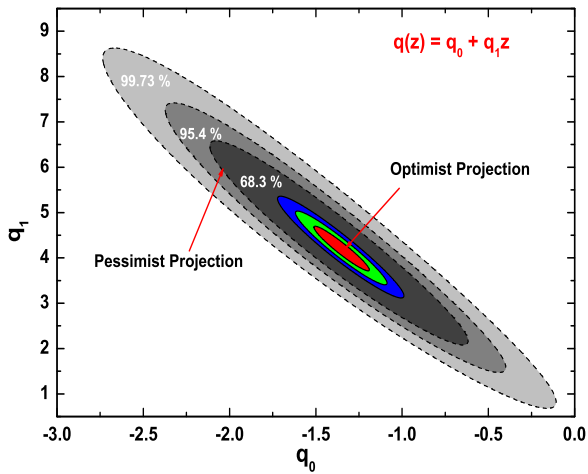,width=3.4truein,height=2.8truein}
\hskip 0.1in} \vspace{-0.5cm} \caption{Contours on the plane
$(q_o,q_1)$ from synthetic Planck data as defined in Table 1. In the
pessimistic case, 210 P (pessimistic projection) and 1050 O
(optimistic projection) clusters were considered by assuming a
random error between 10\% and 20\% in the resulting angular diameter
distances.}
\end{figure}

In the second one (optimistic case - O), 1050 clusters were redshift
distributed as follows, $0\leq z \leq 0.5$ (500), $0.5\leq z \leq 1$
(350)  and $1\leq z \leq 1.5$ (200) with ADD statistical errors  of
$10\%$, $12\%$ and $15\%$, respectively. The redshift intervals were
partitioned into bins ($\Delta z=0.05$) with the clusters
distributed as shown in Table 1\cite{gol}. Both simulations were
carried out by marginalizing over the $H_{0}$ parameter in
$D_{A}(z_i,{\bf p})$ in Eqs. (1) and (5).

In Fig. 3 we display the results of our simulations for the linear
parameterization. The contours correspond $68\%$, $95\%$ and
$99.7\%$ c.l. for the optimistic (colored inner contours) and
pessimistic case (outer contours),  respectively. Comparing with
Fig. 1(a) we see that the allowed region was remarkably reduced even
in the pessimistic case. This means that angular diameters from SZE/X-ray data will
become a potent tool for measuring cosmological parameters fully
independent and competitive with SNe type Ia luminosity distances.

\emph{Conclusions.} We have shown that the combination of
Sunyaev-Zeldovich/X-ray data from galaxy clusters is an interesting
technique for accessing the present accelerating stage of the
Universe.  This result follows from a new kinematic approach based
on the angular diameter distance of galaxy clusters obtained from
SZE/X-ray measurements. By using two different parameterizations, it was found that 
the existence of a transition from a decelerating to an
accelerating expansion occurred relatively recent ($z_{t}\simeq 0.3$).

The ability of the ongoing Planck satellite mission to constrain
the accelerating stage was discussed by simulating two realistic
samples of angular diameters from clusters. The allowed regions in
space parameter was significantly constrained for both the
pessimistic and optimist simulations (Figure 3). The limits on the
transition redshift derived here reinforces the extreme interest  on
the observational search for obtaining SZE/X-ray data from galaxy clusters.

Finally, we  stress  that the present results
depends neither on the validity of general relativity nor the
matter-energy contents of the Universe and, perhaps, more important,
they are  also independent from SNe type Ia observations.

\begin{acknowledgments}
JASL, RFLH and JVC are supported by FAPESP under grants, 04/13668-0,
07/5291-2 and 05/02809-5, respectively. JASL also thanks CNPq (Brazilian Research Agency).
\end{acknowledgments}

\end{document}